\documentclass{article}
\usepackage{spconf,amsmath,graphicx}

\usepackage{soul,color}

\usepackage{hyperref}

\title{Fusion of AUDIO AND VISUAL EMBEDDINGS FOR SOUND EVENT LOCALIZATION AND DETECTION}
\name{Davide Berghi, Peipei Wu, Jinzheng Zhao, Wenwu Wang, Philip J. B. Jackson}
\address{Centre for Vision, Speech and Signal Processing (CVSSP), University of Surrey, Guildford, U.K.}
\begin{document}
%\ninept
%
\maketitle
\begin{abstract}
Sound event localization and detection (SELD) combines two subtasks: sound event detection (SED) and direction of arrival (DOA) estimation. 
SELD is usually tackled as an audio-only problem, but visual information has been recently included.
%Sound event localization and detection (SELD) is usually tackled as an audio-only problem. 
 %thanks to challenges such as DCASE task3 and L3DAS23.
Few audio-visual (AV)-SELD works have been published and most % existing methods 
%simply integrate highly-performant audio systems with visually-extracted spatial information, 
employ vision via
% are limited to
face/object bounding boxes, or human pose keypoints.
In contrast, we explore the integration of audio and visual feature embeddings extracted with pre-trained deep networks. 
For the visual modality, we tested ResNet50 and Inflated 3D ConvNet (I3D). 
Our comparison of AV fusion methods includes the AV-Conformer and Cross-Modal Attentive Fusion (CMAF) model. Our best models outperform the DCASE 2023 Task3 audio-only and AV baselines by a wide margin on the development set of the STARSS23 dataset, making them competitive amongst state-of-the-art results of the AV challenge, without model ensembling, heavy data augmentation, or prediction post-processing.
Such techniques and further pre-training could be applied as next steps to improve performance.

\end{abstract}
\section{Introduction}
\label{sec:intro}

SELD simultaneously predicts the position or DOA of the active sound sources over time, while recognizing their event class \cite{Adavanne:2019:SELDnet}. This task is vital for a variety of applications, such as human-robot interaction, augmented reality, navigation, smart home, and security, to name a few. 
Over time, researchers have been addressing more and more challenges related to the task, including the detection of polyphonic sounds and simultaneous same-class events, dealing with moving targets, and ignoring external interfering sounds.
In 2019, SELD was proposed for the first time for the DCASE challenge Task 3 \cite{Adavanne2019_DCASE}. Recently, the Sony-TAu Realistic Spatial Soundscapes 2023 (STARSS23) dataset \cite{Shimada2023STARSS23AA} was released. In contrast to previous datasets for SELD, STARSS23 also includes 360° video recordings spatially and temporally aligned to the acoustic sound-field captured by the microphone array. This allows exploration of SELD as a multimodal audio-visual problem, we refer to it as audio-visual (AV)-SELD. The two modalities are complementary: 
% and can be beneficial to the task: 
vision provides high spatial accuracy whereas audio can detect occluded objects \cite{zhao:2023:speakertrackSurvey}. A different scenario for AV-SELD was proposed for the L3DAS23 challenge \cite{Guizzo:2022:L3DAS22}, 
% where 
% the visual component only consisted of 
with a single frontal color image % RGB frame 
for each microphone array position. 
% Therefore, it is not possible to leverage the visual temporal dimension and the full field of vision.
Thus, the visual component offered a static narrow view.

In this paper, we tackle AV-SELD by combining audio and visual feature embeddings extracted with pre-trained architectures. We compare two attention-based approaches to fuse the two modalities and demonstrate a substantial improvement over the audio-only and audio-visual DCASE baselines on the STARSS23 dataset.
Code to replicate our models is available on \href{https://github.com/dberghi/AV-SELD}{https://github.com/dberghi/AV-SELD}.% \url{https://github.com/dberghi/AV-SELD}.
%The code to replicate our models is freely available on GitHub\footnote{\url{https://github.com/dberghi/AV-SELD}}.

\section{Related Works}
\label{sec:backgr}

Sound scenes for SELD are usually provided in two spatial acoustic formats: First-Order Ambisonics (FOA), or tetrahedral microphone array (MIC). 
Initially, convolutional recurrent neural networks (CRNN) were the most adopted choice to tackle the task \cite{Adavanne:2019:SELDnet, Cao:2019:polyphonic,Nguyen:2021:SALSA, Wang:2020:dcase20}. However, attention-based architectures, such as Multi-Head Self-Attention (MHSA) \cite{cao:2021:EINv2}, Transformers \cite{Wu:2023:PLDISET}, or Conformers \cite{Gulati2020ConformerCT,Wang:2023:ACS}, are often preferred lately. 
As audio input features, Cao \textit{et al.}~\cite{Cao:2019:polyphonic} proposed to concatenate multichannel log-mel spectrograms with the GCC-PHAT %\cite{Knapp:gccphat:1976} 
computed at each pair of microphones. For the FOA format, intensity vectors (IV) in log-mel space are adopted \cite{cao:2020:EIN}. Nguyen \textit{et al.}~\cite{Nguyen:2021:SALSA} proposed the Spatial Cue-Augmented Log-Spectrogram (SALSA) features, and SALSA-Lite \cite{Nguyen:2021:SALSALiteAF} for the MIC format. To solve the problem of concurrent same-class events, an Event Independent Network \cite{cao:2021:EINv2} with permutation invariant training was proposed. Inspired by this, Shimada \textit{et al.}~\cite{Shimada:2022:multiACCDOA} proposed multiple activity-coupled Cartesian DOA (m-ACCDOA) vectors as SELD output representation. With m-ACCDOA the network is trained with class-wise auxiliary duplicating permutation invariant training (ADPIT).

\begin{figure*}[tb]
\centerline{\includegraphics[width=\textwidth]{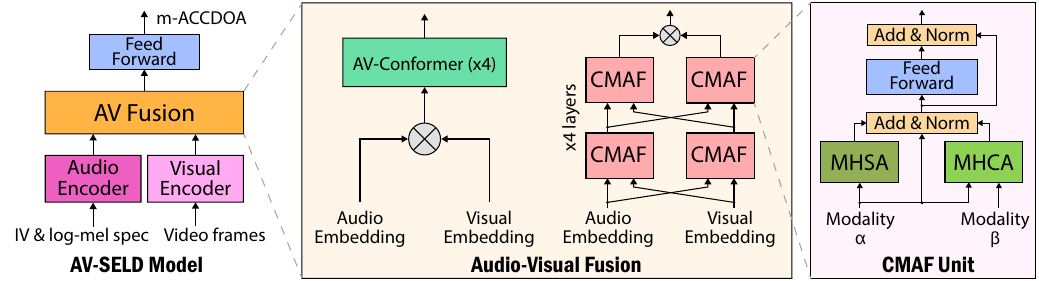}}
\vspace{-1ex}
\caption{(left) AV-SELD pipeline diagram 
with CNN-Conformer audio encoder \cite{Gulati2020ConformerCT}, and I3D \cite{Carreira2017QuoVA} or ResNet-Conformer \cite{He:2016:resnet}\cite{Gulati2020ConformerCT} visual encoder; (center) four layers of attention-based fusion of audio and visual embeddings: AV-Conformer\,\cite{Gulati2020ConformerCT} or Cross-Modal Attentive Fusion, where $\otimes$ is concatenation; (right) CMAF unit diagram \cite{Xinyuan:2023:AVcrossAtt}.}
\vspace{-2ex}
\label{Fig:models}
\end{figure*}

To the best of our knowledge, there is no prior work on AV-SELD before the DCASE 2023 Task 3 challenge, or at least not with a similar setting. Therefore, we will describe the systems proposed by the teams that participated in the challenge, even though the technical reports are too short to provide a thorough description.
%Only a few teams participated in the AV-SELD challenge. 
The AV baseline system consists of a CRNN that leverages visual object detection. Objects bounding boxes are represented as Gaussian-like vectors, as in \cite{Xinyuan:2023:AVcrossAtt}, and concatenated to the features extracted by the audio CNN, before being fed to the recurrent unit. 
Wang \textit{et al.}~\cite{YiWang:2023:dcase23} adopted a similar approach. However, alongside the Gaussian-like vectors, they also included a frame embedding extracted with ResNet18 from the first video frame.  
Wang \textit{et al.}~\cite{Wang:2023:dcase23} performed human keypoint detection and object detection with a ResNet-Conformer model for the AV-SELD predictions. The predictions are then post-processed by taking full advantage of the spatial accuracy of the visual modality. For instance, if the model predicts the class ``Footsteps'', the predicted location is replaced with the location of the closest foot keypoint. %A similar idea has been applied in the \hl{past for AV active speaker localization} 
Kang \textit{et al.}~\cite{Kang:2023:dcase23} adopted a student-teacher distillation to train a ResNet-Conformer from a ResNet-GRU network. For the visual modality, they employed object detection. Kim \textit{et al.}~\cite{Kim:2023:dcase23} trained a network similar to the audio-only baseline, i.e., a CRNN followed by MHSA. In contrast to the other systems, they employed a 3D CNN to extract a visual embedding from raw video, optical flow frames, and object detection features. The so-extracted visual embedding is then concatenated with the audio embedding after the MHSA block, before the prediction of m-ACCDOA vectors, preventing attention-based modality fusion. 

In this work, we extract an audio feature embedding using a CNN-Conformer network and a visual embedding employing either a ResNet-Conformer or the Inflated 3D ConvNet (I3D) \cite{Carreira2017QuoVA}. The two feature embeddings are then fused by an attention-based approach: we compare an AV-Conformer and the Cross-Modal Attentive Fusion (CMAF) network \cite{Xinyuan:2023:AVcrossAtt}.

\section{METHOD}
\label{sec:method}

As depicted in Fig.\,\ref{Fig:models}, our method extracts audio and visual feature embeddings with an audio and a visual encoder, respectively. The embeddings are then combined in an attention-based mechanism. Finally, the output features are fed to two fully connected layers to predict m-ACCDOA vectors for SELD \cite{Shimada:2022:multiACCDOA}. Among our experiments, we studied two different visual encoders and two fusion mechanisms. Below we describe each component and the AV dataset.

%\subsection{Dataset}
\label{ssec:dataset}

\textbf{Dataset:} Results supporting this study are generated on the development set of the STARSS23 dataset \cite{Shimada2023STARSS23AA}. It consists of about 7.5h of real AV spatial recordings of acoustic scenes temporally and spatially annotated with the acoustic class of the active events. It presents a total of 13 event classes. Our experiments employ the FOA audio format. Video recordings consist of 360° views of the environment captured from the microphone array perspective at a resolution of 1920x960p, 30fps. Audio is sampled at 24kHz, 16 bits. Labels are provided at a resolution of 100ms. The dataset includes directional interferes that do not belong to the target classes and must not be detected. It is common to encounter up to 3 simultaneous events, but more simultaneous occurrences, up to 5, are possible.
We employed the provided training-testing split of the development set to train and test our models.

%\subsection{Audio Encoder}

\textbf{Audio Encoder:} The audio encoder takes as input acoustic features extracted from the FOA spatial sound. We employed intensity vectors (IV) in the log-mel space concatenated with the log-mel spectrograms extracted from the FOA channels, yielding 7-dimensional input features with shape $7\times T_{in}\times F_{in}$, where $T_{in}$ corresponds to the number of temporal bins and $F_{in}$ the frequency bins.
The audio encoder includes a CNN architecture followed by a Conformer \cite{Gulati2020ConformerCT}. The CNN presents 4 convolutional blocks with residual connections. Each block consists of two 3$\times$3 convolutional layers followed by average pooling, batch normalization, and ReLu activation. The average pooling layer is applied with a stride of 2, halving the temporal and frequency dimension at each block. The resulting tensor of shape $512\times T_{in}/16\times F_{in}/16$ is then reshaped and frequency average pooling is applied to achieve a $T_{in}/16\times 512$ dimensional feature embedding. $T_{in}$ is chosen so that $T_{in}/16$ matches the label frame rate. %(10 labels per second).
The feature embedding is further processed by a Conformer module with 4 layers and 8 heads each. The size of the kernel for the depthwise convolutions is set to 51.

%\subsection{Visual Encoder}

\textbf{Visual Encoder:} As visual encoders, we compared the ResNet-Conformer and I3D. 
In the former, we fed each video frame to ResNet50 \cite{He:2016:resnet} at a frame rate of 10 fps. In such a way, we extract a number of frame embeddings that match the label frame rate as well as the audio embedding temporal resolution. We then process the frame embeddings with a Conformer module identical to the one employed in the audio encoder.
Additionally, we tested a 3D CNN to take full advantage of the video temporal dimension. We opted for I3D \cite{Carreira2017QuoVA} as it is a widely employed yet effective model for action recognition and 3D feature extraction. The original model comprises two branches: one for the RGB video and one for the optical flow frames. For simplicity, we only employed the RGB branch. 
For computational reasons, we kept the input frame rate adopted with the ResNet-Conformer (10fps). 3D CNNs have the effect of downsampling the temporal dimension. Therefore, the output of I3D is then temporally interpolated to reconstruct the label frame rate. 
The video segments used as inputs to the visual encoders were resized to 448x224p. Both ResNet50 and I3D were pre-trained on squared 224x224p frames. Therefore, to avoid losing spatial information, we divided the input frames into a `left' and a `right' side of 224x224p each, and concatenated the two resulting vectors. In practice, the last layer of ResNet50 is a global average pooling applied on a 7x7 feature map. Since the input horizontal dimension is 448, the resulting feature map has shape 14x7. Therefore we replaced the last global average pooling with a 7x7 average pooling kernel with stride equal to 7, obtaining two output vectors that are then concatenated. 
In both cases, the outputs of the encoders are fed to a linear layer to achieve a 512-dimensional video embedding.

%\subsection{Audio-Visual Fusion}

\textbf{Audio-Visual Fusion:} We considered two attention-based fusion approaches. The first is an AV-Conformer that takes as input the concatenation of the audio and visual embeddings, i.e., a 1024-dimensional vector at each time frame. The AV-Conformer is identical to the one employed in the audio CNN-Conformer and in the ResNet-Conformer. 
The second approach leverages the CMAF network proposed by Qian \textit{et al.} \cite{Xinyuan:2023:AVcrossAtt}. The audio and visual embeddings are the inputs of two parallel CMAF blocks. The structure of each block for two generic modalities $\alpha$ and $\beta$ is depicted in Fig.\,\ref{Fig:models} (right).
In contrast to a traditional Transformer \cite{Vaswani:2017:AttAllUNeed}, in parallel to the MHSA module, they present Multi-Head Cross-Attention (MHCA) where $\alpha$ is used to generate the query $\textbf{Q}_\alpha$ to be mapped to the set of key $\textbf{K}_\beta$ and value $\textbf{V}_\beta$ pairs generated from $\beta$. The input $\alpha$ is then added to the outputs of the MHSA and the MHCA modules.
Two parallel CMAF blocks represent one layer. Their outputs are the inputs for the following layer. %which is also comprised of two parallel CMAF blocks. 
We employed a total of 4 layers. Both MHSA and MHCA contain 8 heads.
The two outputs of the last layer are concatenated to generate a 1024-dimensional vector at each time frame.
We generate $N=3$ tracks as m-ACCDOA vectors via two fully-connected layers. The network is trained with class-wise ADPIT loss \cite{Shimada:2022:multiACCDOA}.

\section{EXPERIMENTS}
\label{sec:experiments}

\begin{figure}[tb]
\centerline{\includegraphics[width=\columnwidth]{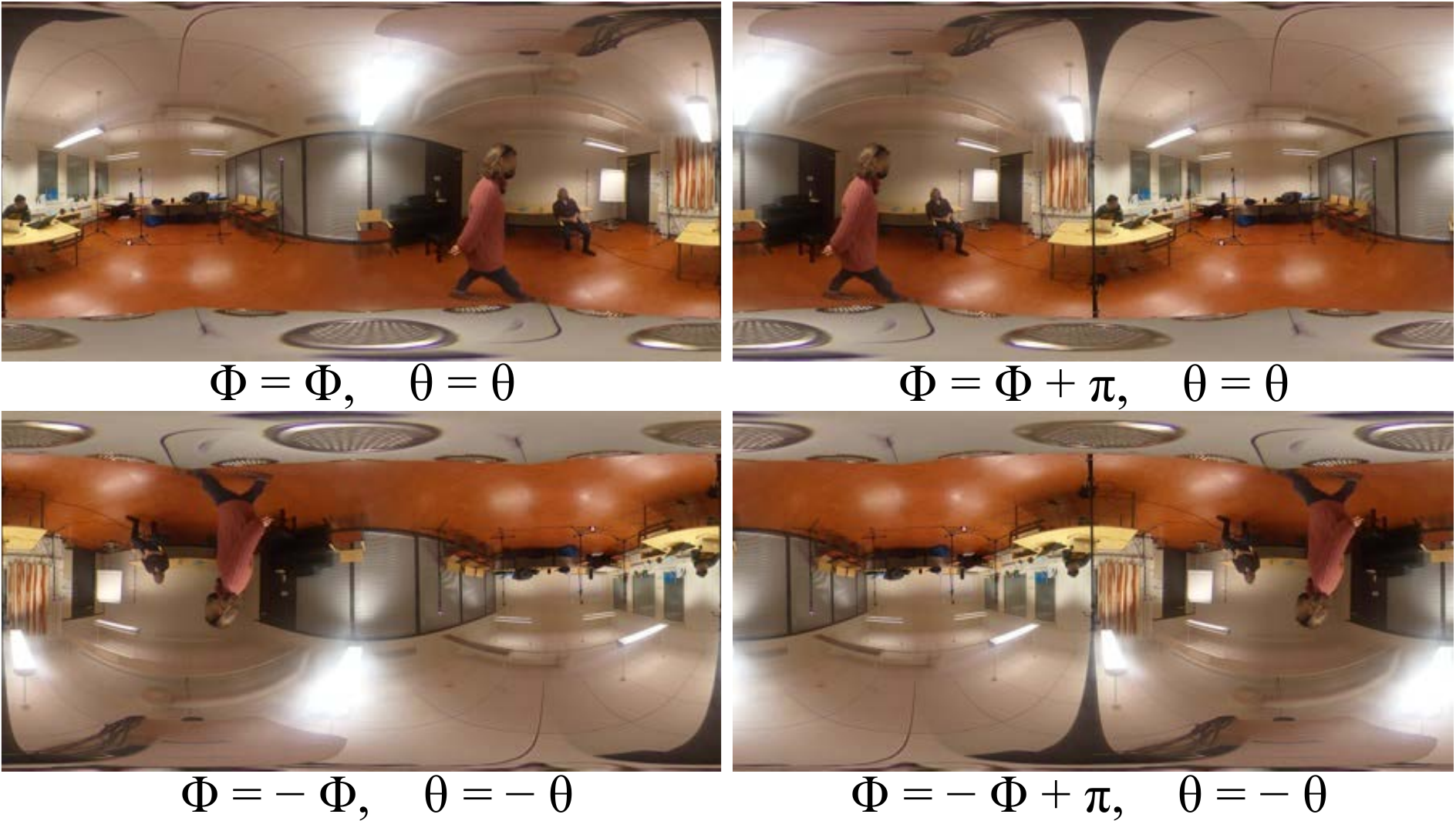}}
\vspace{-1ex}
\caption{Visual transformations for DOA data augmentation.}
\vspace{-2ex}
\label{Fig:AVCS}
\end{figure}

%\subsection{Data Augmentation and Pre-training}
\textbf{Data Augmentation and Pre-Training:} The audio CNN-Conformer was pre-trained on SELD employing the simulated data generator script provided for the DCASE 2022 challenge \cite{Politis2022DCASE}. This allowed the generation of $\sim$30h of spatial recordings, including noiseless and noisy versions. The ResNet50 model employed to process 2D frames is available with the Torchvision library and it is pre-trained on ImageNet. I3D was pre-trained on action detection on the Kinetics dataset.
The synthetic SELD dataset used to pre-train the audio encoder was augmented by a factor of 8 with the audio channels swap (ACS) technique \cite{Wang:2023:ACS}. For the AV-SELD dataset, we also augmented the visual modality consistently with the ACS transformation as indicated by Wang \textit{et al.} \cite{Wang:2023:dcase23}. However, while they only augment the positions of the detected human keypoints or objects bounding boxes, we generate complete new video frames by flipping and rotating the original ones. This creates an effective audio-visual spatial transformation: the audio-visual channel swap (AVCS). Examples of a transformed frame are shown in Fig.\,\ref{Fig:AVCS}.

%\subsection{Implementation Details and Evaluation Metrics}
\label{ssec:metrics}

\textbf{Implementation Details and Evaluation Metrics:} To train our models, we divided the dataset into chunks of 3 seconds, extracted at steps of 0.5s for training and with no overlap for testing.
An STFT with 512-point Hann window and hop size of 150 samples generates spectrograms discretizing the 3-second audio chunks (24kHz) into 480 temporal bins ($T_{in}$). We used 128 frequency bins to generate log-mel spectrograms and intensity vectors (IV) in log-mel space.   
We trained our models with batches of 32 inputs and Adam optimizer for 50 epochs, then we selected the best epoch. The learning rate is set at a fixed value for the first 30 epochs, then and it is decreased by 5\% every epoch. The initial value is in the range of 1E-4 and 1E-3, depending on the one that generates the best results for each model.
The metrics adopted for the evaluation are the ones proposed in the DCASE23 Task3 Challenge \cite{Shimada2023STARSS23AA}. They include location-dependent detection metrics, such as the Error Rate ($ER_{\le20^{\circ}}$) and F1-score ($F1_{\le20^{\circ}}$), computed with respect to a spatial threshold $T=20^{\circ}$, and the Localization Error ($LE_{CD}$) and Localization Recall ($LR_{CD}$). $F1_{\le20^{\circ}}$, $LE_{CD}$, and $LR_{CD}$ are macro-averaged across classes. For compactness, we will omit the subscripts $\le20^{\circ}$ and $CD$ from now on. Results also include the overall $SELD$ score that encodes all other four metrics: $SELD=mean(ER,\,(1-F1),\,LE/180,\,(1-LR)).$

%\begin{equation}
%SELD=mean(ER,\,(1-F1),\,LE/180,\,(1-LR)).
%\label{eq:seld}
%\end{equation}

\begin{table}[tb]
\caption{Results on STARSS23 development set. $F1$ and $LR$ are in percent\,(\%), $LE$ in deg\,(°). CNN-Conf is employed as audio encoder. I3D:\,Inflated 3D ConvNet; RC:\,ResNet-Conformer; Conf:\,AV-Conformer; CMAF:\,Cross-Modal Attentive Fusion; CA:\,Cross-Attention; GRU:\,Gated Recurrent Units; AO:\,Audio-only; AV:\,audio-visual.}
\vspace{-1ex}
\begin{center}
\begin{tabular}{c|c|c|c|c|c|c}
\hline
\textbf{Visual}&\textbf{Fusion}&\textbf{ER$\downarrow$}&\textbf{F1$\uparrow$}&\textbf{LE$\downarrow$}&\textbf{LR$\uparrow$}&\textbf{SELD$\downarrow$}\\  \hline
I3D & CMAF & 0.57 & 40.5 & 33.2 & 55.3 & 0.45 \\
%\hline
I3D & Conf & 0.52 & 46.4 & 16.9 & 60.2 & 0.39 \\
%\hline
RC & CMAF & 0.54 & 41.3 & 31.6 & 53.4 & 0.44 \\
%\hline
RC & Conf & 0.51 & 49.5 & 15.8 & 60.2 & \textbf{0.38} \\
%\hline 
RC & CA & 0.55 & 34.7 & 30.4 & 47.8 & 0.47 \\
%\hline
RC & GRU & \textbf{0.50} & 49.4 & 16.2 & 56.8 & \textbf{0.38} \\
%\hline
Both & Conf & 0.52 & 48.0 & 16.2 & \textbf{60.8} & \textbf{0.38} \\
\hline 
\hline
\multicolumn{2}{c|}{Visual-only} & 1.03 & 0.9 & 103 & 11.4 & 0.87 \\ 
%\hline
\multicolumn{2}{c|}{Audio-only} & 0.51 & \textbf{50.2} & \textbf{15.4} & 56.4 & \textbf{0.38} \\ 
\hline 
\hline
\multicolumn{2}{c|}{Baseline AO} & 0.57 & 29.9 & 22.0 & 47.7 & 0.48 \\ 
%\hline
\multicolumn{2}{c|}{Baseline AV} & 1.07 & 14.3 & 48.0 & 35.5 & 0.71 \\  \hline

\end{tabular}
\vspace{-4ex}
\label{tab:results}
\end{center}
\end{table}

%\subsection{Experiments}
\label{ssec:experiments}

\textbf{Experiments:} In addition to the combinations of ResNet-Conformer/I3D and CMAF/AV-Conformer introduced previously, we tested a simplified cross-attention (CA) mechanism that combines the audio and visual embeddings like in CMAF. However, each CA block only presents MHCA without parallel self-attention.
We also compared our attention-based fusion methods with a recurrent network, i.e., two bidirectional gated recurrent units that take as input the concatenation of the embeddings. 
Furthermore, we conducted an experiment where both I3D and ResNet-Conformer visual embeddings are concatenated to the audio embedding before being fed to the AV-Conformer.
Results are reported in Tab.\,\ref{tab:results}. The table also includes visual-only (VO) and audio-only (AO) methods to understand the contribution of each modality. For fairness, we matched the networks' depth of the AV methods by adding an additional 4-layer Conformer to the CNN-Conformer and the ResNet-Conformer, respectively.
%Therefore, AO and VO respectively consist of the CNN-Conformer and the ResNet-Conformer followed by an additional 4-layer Conformer.
The bottom rows report the results achieved by the AO and AV challenge baselines.

%\subsection{Results}
\label{ssec:results}

\textbf{Results:} Comparing the single-modal methods to the AV methods, it is clear that most of the task is accomplished by the audio modality, which alone got the best $LE$ and $F1$. Nevertheless, results from VO suggest that the visual modality can provide a partial contribution to that task, especially in terms of $LR$ which is above 11\%. In fact, the AV methods that employ the AV-Conformer fusion achieved up to a 4 percentage points gain in $LR$ over the AO system ($LR>60\,\%$).
However, vision did not contribute to improving the $ER$ and spatial accuracy.
Comparing I3D and ResNet-Conformer, it is possible to appreciate a slight benefit with the latter, even though it is marginal. Nevertheless, the method that leverages both visual encoders is the one with the highest gain in $LR$. %It achieves a class-independent \textit{micro}-averaged $LE$ of 18.6{°}.

While CMAF proved to be preferable over direct cross-attention, the methods trained with this type of AV fusion produced weak results compared to the AV-Conformer or GRU, with $LE>30^{\circ}$. From a thorough analysis, we noticed that the problem is caused by the `Knock' class, which CMAF and CA are not detecting. This particularly penalizes the $F1$, $LE$, and $LR$ as they are macro-averaged across classes. For CMAF with ResNet-Conformer as visual encoder, we re-computed the micro-averaged metrics and got $F1=54.3\%$, $LE=19.0^{\circ}$, and $LR=71.4\%$. 
For comparison, the system that employs the AV-Conformer and both visual encoders achieves micro-averaged $F1=55.4\%$, $LE=18.6^{\circ}$, and $LR=72.6\%$.
This suggests that CMAF and CA are more competitive than what appears in the table. While it is not a justification for not detecting the `Knock' class, it should be noted that `Knock' appears in only 90 frames out of the entire training set (9 seconds). We argue that results might be different with a more balanced dataset.

AV-SELD is a new and non-trivial task. This can be appreciated from the weak results achieved by the AV baseline. Yet, all our methods outperformed the baselines by a wide margin. To our knowledge, only Wang \textit{et al.} \cite{Wang:2023:dcase23} and Kang \textit{et al.} \cite{Kang:2023:dcase23}, the challenge top-2 teams, achieved higher performance on the AV-SELD task. We believe that integrating visually-guided prediction post-processing, as in \cite{Wang:2023:dcase23}, could help improve DOA estimation of our systems alongside $ER$ and $F1$. Further improvement can be achieved from additional pre-training, data augmentation, and model ensembling.

\section{CONCLUSION}
\label{sec:conclusion}
In this paper, we proposed and compared some of the first architectures for AV-SELD. We tackled the task by combining audio and visual feature embeddings extracted with pre-trained deep networks, comparing two well-established visual encoders. In terms of AV fusion, tests include but are not limited to, the AV-Conformer and CMAF. Our methods substantially outperformed the audio-visual and audio-only baselines of the DCASE 2023 Task 3 challenge.

%\vfill\pagebreak

\section{Acknowledgments}
\label{sec:Ack}

Research supported by BBC Prosperity Partnership AI4ME: Future Personalised Object-Based Media Experiences Delivered at Scale Anywhere EP/V038087. For the purpose of open access, the author has applied a Creative Commons Attribution (CC BY) licence to any Author Accepted Manuscript version arising. Data supporting this study available from \url{https://zenodo.org/record/7880637}.

\bibliographystyle{IEEEbib}
\bibliography{refs}

\end{document}